# Resampling Residuals on Phylogenetic Trees: Extended Results

**Peter J. Waddell[1], Ariful Azad[2] and Ishita Khan[2]**

**pwaddell@purdue.edu**

[1]Department of Biological Sciences, Purdue University, West Lafayette, IN 47906, U.S.A.

[2]Department of Computer Science, Purdue University, West Lafayette, IN 47906, U.S.A

.

In this article the results of Waddell and Azad (2009) are extended. In particular, the geometric percentage mean standard deviation measure of the fit of distances to a phylogenetic tree are adjusted for the number of parameters fitted on the tree. The formulae are also presented in their general form for any weight that is a function of the distance. The cell line gene expression data set of Ross et al. (2000) is reanalyzed. It is shown that ordinary least squares (OLS) is a much better fit to the data than a Neighbor Joining or BME tree. Residual resampling shows that cancer cell lines do indeed fit a tree fairly well and that the tree does have strong internal structure. Simulations show that least squares tree building methods, including OLS, are strong competitors with BME type methods for fitting model data, while real world examples often suggest the same conclusion.

"… his ignorance and almost doe-like naivety is keeping his mind receptive to a possible solution." A quotation from Kryten: Red Dwarf VIII-Cassandra

**Keywords**: Resampled Residual Bootstrap, flexi-Weighted Least Squares Phylogenetic Trees fWLS, Balanced Minimum Evolution BME, Phylogenomics, Gene Expression Tree

## 1 Introduction

This article updates and extends some of the results in Waddell and Azad (2010). In particular the fit equations of the form %SD and g%SD from Waddell et al. (2007) are presented in their general form for any function of the distances used as a weighting function in least squares fitting. The effect of this change on the plots of Waddell and Azad (2009) is discussed and illustrated.

A much larger data set, based on the gene expression profiles of 64 NCI cancer cell lines (Ross et al. 2000), is analyzed using the methods of Waddell and Azad (2009). This is an interesting data set since it was one of the first to show that cancers can be clustered based on their gene expression profiles. In doing so, it revealed cryptic forms of cancer; that is, the histology of the cancer looked similar but the gene expression profile suggested it was quite a different disease.

The use of NJ and BME with residual resampling is further explored and simulations are used to more clearly define which distance-based method(s) are best for analyzing the Ross et al. (2000) data.

## 2 Materials and Methods

The materials and methods follow directly from Waddell and Azad (2009). The analyses in this article use the latest version of PAUP*4.0 a114 (Swofford 2000). The distances between cell lines were estimated by downloading the Ross et al. (2000) data, selecting the 1161 genes they used in figure 1 and using Microsoft Excel to estimate the Pearson correlation between the expression profiles of each pair of cell lines, in order to derive a distance of 1 – (the pairwise correlation). This is the same distance function used by Ross et al. (2000) in their figure 1.

## 3 Results

The results below extend the results presented in Waddell and Azad (2009). Section 3.1 presents the percentage fit equations in a general form, where the weights on the squared residuals are any function of the observed or the expected distance.

### 3.1 Flexi-Weighted Least Squares Criteria

The fit equations in Waddell, Kishino and Ota (2007) and Waddell and Azad (2009) can be reexpresssed in a general form, based on the function of the distances that is being used for the weights. The weights can be considered a proxy for the variance of the squared errors.

When the weights are of the form $\sigma_i^2 = \sigma^2 d_i^P = c d_i^P$,

$$\%SD = \left(\frac{1}{N}\sum_{i=1}^{N} d_{obs_i}\right)^{-1} \left(\frac{1}{N}\sum_{i=1}^{N} d_i\right)^{P/2} \left(\frac{1}{N}\sum_{i=1}^{N} \frac{(d_{obs_i} - \hat{d}_{\exp_i})^2}{d_i^P}\right)^{0.5} \times 100\% \qquad \text{(eq 1)}$$

where $d_i$ can be either the observed or the expected distance.
More generally, we can envisage the form

$$\%SD = \left(\frac{1}{N}\sum_{i=1}^{N} d_{obs_i}\right)^{-1} \left(f\left(\frac{1}{N}\sum_{i=1}^{N} d_i\right)\right)^{0.5} \left(\frac{1}{N}\sum_{i=1}^{N} \frac{(d_{obs_i} - \hat{d}_{\exp_i})^2}{f(d_i)}\right)^{0.5} \times 100\% \qquad \text{(eq 2)}$$

for those cases where the variance is proportional to any function of the distance.

If the model of the variance is $\sigma_i^2 = \sigma^2 w_i = c w_i$, then we have the general scale-free fit monotonic with likelihood being,

$$g\%SD = \left(\prod_{i=1}^{N} d_{obs_i}\right)^{-1/N} \left(\prod_{i=1}^{N} w_i\right)^{\frac{1}{2N}} \left(\frac{1}{N}\sum_{i=1}^{N}\frac{(d_{obs_i} - \hat{d}_{\exp_i})^2}{w_i}\right)^{0.5} \times 100\% \qquad \text{(eq 3)}$$

which can also be expressed as,

$$g\%SD = \left(\prod_{i=1}^{N} \frac{w_i}{d_{obs_i}^2}\right)^{\frac{1}{2N}} \left(\frac{1}{N}\sum_{i=1}^{N}\frac{(d_{obs_i} - \hat{d}_{\exp_i})^2}{w_i}\right)^{0.5} \times 100\% \qquad \text{(eq 4)}$$

Above in equations (3) and (4) are the likelihood inspired versions of the geometric mean fit percent deviation fit statistic, which is inversely monotonic with the likelihood. However, since these models fit also fit into the framework of general linear models, these statistics can be based on unbiased estimators of the variance. In this case replace the normalization term $1/N$ immediately before the sum of squares with the term $1/(N\text{-}k)$, where $N$ is the number of pieces of information (the number of unique distances) while $k$ is the number of free parameters estimated by the model (e.g., Agresti 1990). Note, this property of unbiasedness only strictly holds when the generating tree is known and it is allowed to have negative edge lengths. Otherwise, the true expected values of the variance deviate due to tree selection, constraint of edge lengths to be non-negative and constraint that $d_{\exp}$ be non-negative (the latter being a weaker condition than the former).

**3.2 Reevaluation of the Yeast results with 1/(*N* – *k*)**

As mentioned above, the fit statistics can be derived using a less biased estimate of the variance of the data. In general linear Gaussian models, each parameter fitted away from a boundary is expected to reduce the sum of squares by 1 unit. Thus if k parameters are fitted, then these are expected to reduce the total sum of squares to $N - k$, where $N$ is the number of pieces of information, in this case, the pair-wise distances. This holds exactly on the ordinary least squares tree used to generate the data, where edges are allowed to be negative. If edges are constrained to be non-negative, then this can initially increase the total sum of squares beyond that expected due to prohibiting negative edge lengths that might improve the fit. Tree search, which is model selection, adds a whole new level of complexity as to what to expect. In general, it will drive down the sum of squares more than what is expected by the factor $k$. The easiest way to ascertain this is using simulations. What exactly occurs when edge lengths are constrained to be non-negative and there is tree search is a balance of two different effects. As long as internal edges in the tree are a few times as large as the errors expected on the distances traversing them, then there tends to be little impact of the non-negative boundary and few other trees close to the optimal positively weighted generating tree, then empirical sum of squares multiplied by $1/(N - k)$ tends to be a reliable and close to unbiased biased estimator of the true variance (simulations not shown).

With this in mind we replot two of the figures from Waddell and Azad (2009) in figure 1. Note that when $1/N$ is replaced by $1/(N-k)$ all the g%SD values change by a factor of $\sqrt{\frac{N}{N-k}}$, thus they get uniformly bigger. This does not change the minimum for the g%SD curve, nor does it change the ML point. It is clear from figure 1 that the real protein based sequences are fitting better than the DNA based sequences. The implied g%SD error has also increased considerably, since $N$ is small (28) and $k$ is relatively large ($2t - 3 = 13$), so the factor $\sqrt{\frac{28}{28-13}} = 1.366$ is an appreciable difference.

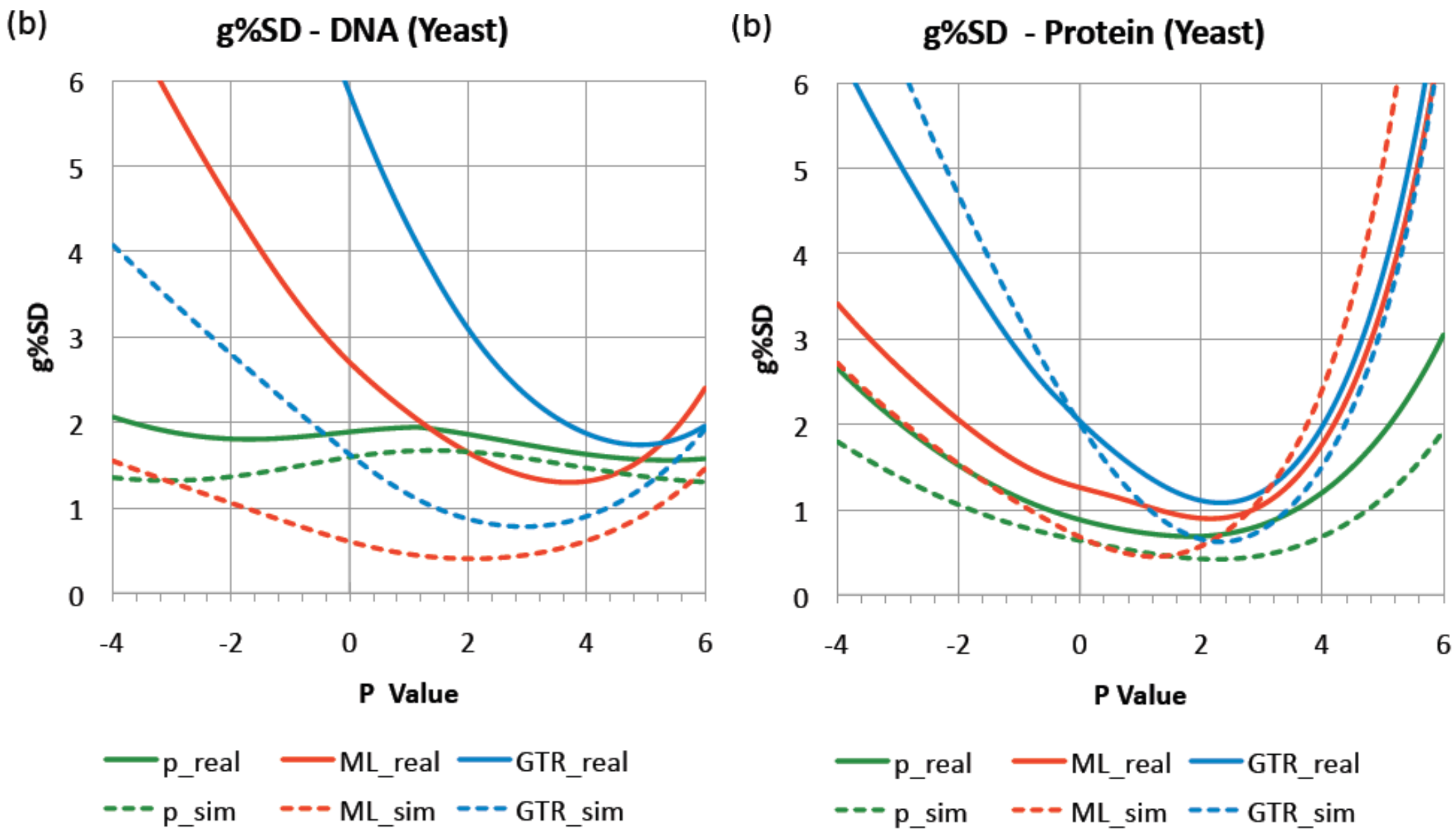

Figure 1. The g%SD fit figures for fWLS using polynomial weights of the form $d_{obs}^{P}$, fitted to yeast genomic sequences, and the a less biased variance estimator of the form $1/(N-k)$. (a) DNA sequences (b) Protein sequences.

### 3.3 fWLS tree of cancer cell lines

Reanalyzing the distance data of Ross et al. (2000) with polynomial weights, we found the maximum likelihood and minimum g%SD occurred at P = 0 to -0.3 with g%SD = 7.590. The absolute best fit was near -0.2. This indicates that the average error of fitting the distances to the tree is nearly independent of the size of the distance. For simplicity and to exploit the faster search algorithms available for P = 0 in PAUP*, we set P to 0.

A surprising feature of this data is how well the data fits a tree despite there being no strong *a priori* argument that it must do so. Indeed, the fit observed here with a mean per cent standard deviation of 7.480 and a g%SD of 7.590 is considerably better than commonly seen with morphological phylogenetic data sets (unpublished analyses). We hypothesize that this fit is due to a tree-like progression of epigenetic processes, perhaps going back to the original stem cells, and remembered by both the original tissue, the cancer and the cell line(s) established from the cancer. If this is the case, then it strongly reinforces the arguments for fitting such tissue expression data to a tree.

The optimal tree is shown in figure 1. It is slightly non-clock like and it differs from the unweighted pair group method dendrogram (tree) of Ross et al. (2000) mostly in the deeper parts of the tree. Surprisingly however, these parts of the tree include some very supported groups when the residuals are resampled, added to the tree based distances to produce 1000 pseudo-replicate data sets, a tree reestimated independently on each data, and the majority-rule consensus tree of the 1000 trees constructed. These include multiple support values in the range of 99 to 100%. Lower down in the tree, all the large monotypic clusters of cancer types are supported with values of greater than 95%. Note that the breast cancers (pink) amongst the melanoma cancers (brown) are now accepted as being an occult or undetected metastasis of a melanoma (Ross et al. 2000). The data not only fit a tree well, but the tree is also showing a good deal of robust hierarchical structure.

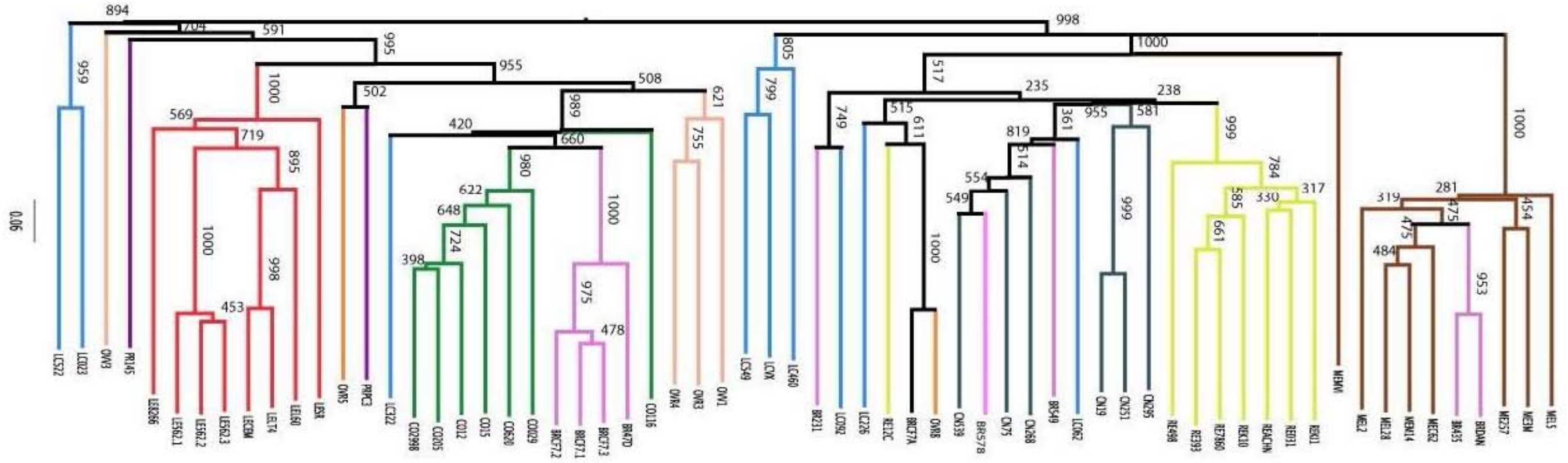

Figure 1. The unweighted least squares tree constrained to non-negative edge lengths for distances calculated from the expression profiles of 1161 genes used by Ross et al. 2000 for their figure 1. The distance was recalculated as 1- the Pearson correlation of the expression profile of each pair of cell lines. The %SD is 7.480 and the g%SD is 7.590. The numbers associated with each edge are how often that edge was recovered when re-adding the residuals to the expected or tree distance. It was iterated once since re-adding the residuals followed by tree search reduced the mean g%SD of the replicates relative to the original data by ~1%.

### 3.4 The performance of NJ and BME in comparison to OLS

The analyses of the previous section were repeated using Neighbor Joining (NJ). Despite NJ recovering a similar tree except for the deeper parts of the tree, the g%SD was very poor at 25.607%. We then applied residual resampling to the tree, and the consensus tree was very poorly resolved. This raised the possibility that residual resampling may not be working with this method of weighted least squares. To further check this possibility, an extensive search for the best BME tree using a fast SPR (Hordijk and Gascuel 2005) and TBR search algorithm in a new test version of the program fastME (Desper and Gascuel 2002, Gascuel Pers Com.). It a found considerably better BME trees, with a g%SD reduced to 19.933%. However, residual resampling about this tree followed by BME with repeated full cycles of tree bisection and rearrangment search also failed to yield a robust consensus tree. A potential reason for this poor result is that BME is assuming errors proportional to $\sqrt{2}$ when deep distances through the tree go between sister tips (thus separated by one internal node), but assuming similar distances through the tree that traversed many internal nodes to have an error rate of $\sim\sqrt{500{,}000}$, despite the actual fitting of residuals indicating no evidence for this!

In order to further explore this interesting finding, extensive simulations using BME and OLS weighted trees, were performed. Simulating with the recovered BME tree and that models errors implied by the original data, all the tree reconstruction methods got the vast majority of the tree wrong as seen in table 1. Indeed the proportion of the tree wrong was estimated at ~110/(2(2*t*-3)) = 110/122 or 90.2% or worse of all internal edges. Given that the expected Folds-Robinson distance of random binary trees with 64 tips is ~120, this is a very poor result. Thus, the resampling results obtained with NJ and BME for the Ross et al. data do make sense under the model.

Exploring further, table 1 offers some important insights into the use of distance methods for reconstructing trees with a shape and errors like those of the Ross et al. data. The simulations progressively reduced the variance in steps of a factor of 4 and therefore reduced the g%SD in steps of a factor of two. The best tree building method varied from BioNJ when the variance matched that of the original data, then switched to OLS|| (ordinary least squares with negative edge weights switched to absolute values prior to taking the sum of squares) with variances of ¼ and 1/16. The model used to infer the errors, BME, did not do best at reconstructing the tree until the variance had dropped by a factor of 1/64 and 1/256. This suggests that in cases of large errors, BME does not go especially well even when the errors follow the WLS method used to construct its penalty function. One would expect the BME_WLS+ model to be a best linear unbiased estimator in these circumstances, but it has yet to be implemented. This may be showing up some of the expected weaknesses in the performance of minimum evolution criteria (Gascuel et al. 2001) compared to minimum least squares suggested in Mihaescu and Pachter (2008).

In Table 2 the simulations are upon the OLS+ tree recovered from the Ross et al. distance data with OLS errors matching those seen with the real data (variance = 1). In this situation, for both the original variance and every reduction of it, the method OLS+_TBR consistently outperformed all the other tree building methods. This is consistent with its position as a best linear unbiased estimator under these conditions. At lower variances (1/4 or lower), OLS|| became the second best method and converged towards OLS+ with decreasing variance. The best of the remaining methods tended to be UNJ or unweighted Neighbor Joining (Gascuel 1997b) followed by BioNJ (Gascuel 1997a), which is consistent with the conditions UNJ was designed to perform well in. It is notable how poorly the older version of fastME with BME and only nearest neighbor interchanges (BME_NNI) performed. This method struggled except at the lowest variances, suggesting that the NNI moves used often saw it trapped in local optima. Fortunately, the authors have now devised very fast extensive search methods for BME that perform much better.

These simulation results suggest that both the g%SD and the residual resampling results reported here are reasonable. It also highlights that this does not seem to be a good data set to analyze with BME or related methods like NJ, particularly in terms of resolving the deeper parts of the tree. These simulations show a stronger performance of constrained OLS methods compared to ME and BME type methods than might be gathered from published simulations, even under the BME type model! The weighted tree will clearly have something to do with this. The Ross et al. distances suggest a tree that has markedly longer external edges than is expected in coalescent or Yule-Harding trees. However, such a pattern of longer external edges is quite typical of the phylogenetic trees reconstructed when a few divergent taxa are sampled from each group of interest (as is often the case for the hardest phylogenetic problems). The simulation results also suggest that likelihood and the g%SD measure should be used to suggest which distance methods really meet the data's errors. If this is the lesson, then it parallels what many of us learned at kindergarten; no matter how big the hammer, do not drive the square peg into the round hole (or do not use BME if OLS fits much better)!

Table 1. Simulations with independent normally distributed weighted least squares errors. The weighting on the errors follows the BME model fitted to the Ross et al. (2000) distance data. The g%SD shown is the expected value based on resampling of these errors on the generating tree and reconstruction with the matching WLS method. The variance is the inverse ratio of the simulated variance to that of the original data. The methods of reconstruction used are, in order, NJ in PHYLIP, then in fastBME, NJ, BioNJ, Unweighted NJ, Greedy BME, and Greedy OLS algorithms followed by BME_NNI, BME_SPR and BME_TBR, ME_OLS_SPR and ME_OLS_TBR. Finally, PAUP* was used to infer ordinary least squares trees with OLS_TBR, OLS||_TBR and OLS+_TBR. The numbers shown for each variance and each method are the mean Folds-Robinson unweighted distance from the generating tree.

| G%SD | 19.933 | 9.966 | 4.983 | 2.491 | 1.245 |
|---|---|---|---|---|---|
| Variance | 1 | 4 | 16 | 64 | 256 |
| Method | | | | | |
| NJ_PHYLIP | 110.254 | 96.288 | 76.612 | 46.348 | 1.656 |
| NJ | 110.254 | 96.288 | 76.612 | 46.348 | 1.656 |
| BioNJ | **109.280** | 95.874 | 76.876 | 46.726 | 1.792 |
| UNJ | 113.692 | 97.648 | 76.898 | 46.432 | 1.684 |
| GBME | 114.328 | 94.968 | 69.457 | 26.544 | 0.596 |
| GOLS | 115.014 | 94.872 | 66.360 | 18.082 | 1.356 |
| BME_NNI | 110.548 | 96.392 | 76.348 | 46.004 | 0.596 |
| BME_SPR | 113.208 | 92.266 | 65.932 | 14.578 | **0.038** |
| BME_TBR | 113.192 | 92.018 | 65.470 | **13.994** | **0.038** |
| ME-OLS_SPR | 113.212 | 92.32 | 66.368 | 16.478 | **0.038** |
| ME-OLS_TBR | 113.14 | 92.228 | 66.216 | 16.442 | **0.038** |
| OLS_TBR | 114.884 | 93.036 | 63.158 | 20.670 | 1.198 |
| OLS||_TBR | 113.306 | **91.126** | **58.434** | 18.086 | 1.968 |
| OLS+_TBR | Incomplete | | | | |

Table 2. Simulations with independent normally distributed unweighted least squares errors. The weighting on the errors follows the OLS+ model fitted to the Ross et al. (2000) distance data. The g%SD shown is the expected value based on resampling of these errors on the generating tree and reconstruction with unconstrained OLS.

| g%SD | 7.5924 | 3.7962 | 1.8981 | 0.94905 | 0.4745 |
|---|---|---|---|---|---|
| Variance | 1 | 1/4 | 1/16 | 1/64 | 1/256 |
| Method | | | | | |
| NJ_PHYLIP | 47.460 | 20.524 | 8.114 | 2.422 | 0.234 |
| NJ | 47.460 | 20.524 | 8.114 | 2.422 | 0.234 |
| BioNJ | 45.608 | 18.756 | 6.830 | 1.646 | 0.128 |
| UNJ | 45.122 | 18.276 | 6.538 | 1.492 | 0.124 |
| GBME | 60.110 | 29.260 | 11.886 | 4.288 | 1.028 |
| GOLS | 60.044 | 29.500 | 12.234 | 4.404 | 1.006 |
| BME_NNI | 60.110 | 29.260 | 11.886 | 4.288 | 1.028 |
| BME_SPR | 49.068 | 22.304 | 8.974 | 2.768 | 0.304 |
| BME_TBR | 48.946 | 22.298 | 8.974 | 2.766 | 0.304 |
| ME-OLS_SPR | 48.920 | 22.214 | 8.938 | 2.754 | 0.304 |
| ME-OLS_TBR | 48.892 | 22.270 | 8.964 | 2.760 | 0.304 |
| OLS_TBR | 68.182 | 36.604 | 15.766 | 5.962 | 1.142 |
| OLS\|\|_TBR | 51.136 | 17.818 | 4.940 | 1.038 | **0.050** |
| OLS+_TBR | **37.942** | **14.890** | **4.772** | **1.026** | 0.052 |

## 4 Discussion

The reanalysis of the Ross et al. (2000) data shows that residual resampling and flexi-Weighted Least Squares are useful methods for the analysis of cancer/tissue types. The data are surprisingly tree-like suggesting that epigenetic factors may induce a tree-like structure in the expression profiles of tissues/cancers and cell lines. The residual resampling shows that many parts of the tree of cancer cell lines are robust, including deeper parts of the tree that may disagree with the tree reconstructed by NJ or UPGMA (the latter being used in Ross et al. 2000).

Simulations show that the best method at reconstructing the tree can be very much data dependent. This suggests that rather than blanked recommendations such as use NJ or use fast BME, it is necessary to evaluate the likelihood or g%SD of the data. The later measure has the advantage that biologists can compare it across data sets to hopefully get a better feeling for the potential impact of both sampling and systematic errors upon their inferences. More generally, ordinary least squares fitting may strongly outperform BME, suggesting that fast algorithms for its use in tree search on large data sets should be extended and implemented (Bryant and Waddell 1997, 1998).

"… and with the ignorance he's got that makes him one of the most powerful men that have ever lived." A quotation from Kryten: Red Dwarf VIII-Cassandra

## Acknowledgements

This work was supported by NIH grant 5R01LM008626 to PJW. Thanks to Joe Felsenstein, Olivier Gascuel, Mike Hendy, Hiro Kishino, Mike Steel, and Dave Swofford for helpful discussions. Special thanks to Dave Swofford for use of a test version of PAUP that offers any real value for parameter *P* when using its least squares functions, and Alex Pothen for allowing AA to share time on this project while working on another project.

## Author contributions

PJW originated the research, developed methods, gathered data, interpreted analyses, prepared figures and wrote the manuscript. AA implemented methods in C and PERL, ran analyses, prepared figures, interpreted analyses and commented on the manuscript.